\documentclass[aps,twocolumn,amsmath,amssymb,superscriptaddress,notitlepage]{revtex4-1}
\usepackage{graphics}
\usepackage{bm}
\usepackage{dcolumn}
\usepackage{natbib}
\usepackage{epstopdf}
\usepackage{float}
\usepackage{graphicx}
\usepackage{epsfig}
\usepackage[pdfstartview=FitH]{hyperref}
\usepackage{color}
\usepackage{appendix}
\usepackage{ulem}

\newcommand{\rr}{{\bf r}}

\begin{document}
\title{Chern-number matrix of the non-Abelian spin-singlet fractional quantum Hall effect}
\author{Tian-Sheng Zeng}
\affiliation{Department of Physics, College of Physical Science and Technology, Xiamen University, Xiamen 361005, China}
%\affiliation{School of Science, Westlake University, Hangzhou 310024, China}
\author{W. Zhu}
\affiliation{School of Science, Westlake University, Hangzhou 310024, China}
\date{\today}
\begin{abstract}
While the internal structure of Abelian topological order is well understood, how to
characterize the non-Abelian topological order is an outstanding issue. We propose a distinctive scheme based on the many-body Chern number matrix to characterize
non-Abelian multicomponent fractional quantum Hall states.
As a concrete example, we study the many-body ground state of two-component bosons at the filling faction $\nu=4/3$ in topological flat band models.
Utilizing density-matrix renormalization group and exact diagonalization calculations,
we demonstrate the emergence of non-Abelian spin-singlet fractional quantum Hall effect under three-body interaction,
whose topological nature is classified by six-fold degenerate ground states and fractionally quantized Chern number matrix.
%At generic fillings $\nu=2k/3$, we propose the topological characterization of non-Abelian spin-singlet fractional quantum Hall effects by a universal Chern number matrix.
\end{abstract}
\maketitle

\section{Introduction}

Topologically ordered quantum states like fractional quantum Hall (FQH) state, which are beyond Landau symmetry breaking paradigm,
demonstrate many remarkable features such as fractionalization of quasiparticles and topological ground state degeneracy~\cite{Wen1990a,Wen1990b}. For a FQH state (irrespective of whether it is Abelian or non-Abelian in nature) at given filling $\nu$, it is common knowledge that the Hall conductance should be a quantized value $\sigma_H=\nu$, which is related to a topological invariant (so called Chern number)~\cite{Kohmoto1985}.

As sparked by Laughlin's seminal work~\cite{Laughlin1983}, the Hall conductance is also related to the correlation power $(z_i-z_j)^m$ in the trial wave function $\psi\prod_{i<j}(z_i-z_j)^m$ through the equality $\sigma_H=\nu=1/m$. Since then, trial wave functions have become one of the central roles in the study of spin-polarized FQH effects, like composite-fermion wave functions of Abelian FQH states at filling $\nu=p/(2p\pm1)$~\cite{Jain1989}. In addition, one of the earliest non-Abelian paired states at filling $\nu=1/m$ (odd $m=2$ for fermion and even $m=1$ for bosons) was described by using either a parton wave function~\cite{Wen1991} or a Moore-Read Pfaffian wave function~\cite{Moore1991} $\Psi_{MR}=\text{Pf}(\frac{1}{z_i-z_j})\prod_{i<j}(z_i-z_j)^m$ whose particle-hole symmetric version may be the most promising candidate for the mysterious $\nu=5/2$ FQH effect in experiment~\cite{Dutta2022}. As an extension of the Moore-Read states, a class of single-component non-Abelian parafermionic states at filling $\nu=k/(kM+2)$ (odd $M$ for fermion and even $M$ for bosons) can be constructed as Read-Rezayi clustered wave function~\cite{Read1999} $\Psi_{RR}\propto\prod_{i<j}(z_i-z_j)^{M+2/k}$ apart from a constant conformal correlator. For these spin-polarized states (Laughlin, Moore-Read or Read-Rezayi states), the Hall conductance $\sigma_H=\nu=1/\eta$ is given by the correlation power of $(z_i-z_j)^{\eta}$ in the wave function.

Further, for a wide class of quantum states dubbed as Abelian topological ordered phases where the exchange statistics of quasiparticle belongs to a U(1) group,
it has been recognized that the integer-valued symmetric $\mathbf{K}$ matrix
is capable of classifying the internal topological structure of multicomponent systems at generic fillings~\cite{Wen1992a,Wen1992b,Blok1990,Blok1991}.
The simplest example is two-component quantum Hall state described by the $\mathbf{K}=\begin{pmatrix}
m & n\\
n & m\\
\end{pmatrix}$ matrix ($m,n \in Z$), with the associated Halperin's wave function~\cite{Halperin1983}
\begin{align}
 \Psi_{mmn} \propto \prod_{i<j}(z_i^{\uparrow}-z_j^{\uparrow})^{m} (w_i^{\downarrow}-w_j^{\downarrow})^{m} \prod_{i,j}(z_i^{\uparrow}-w_j^{\downarrow})^{n}, \nonumber
\end{align}
where $z_i^{\uparrow},w_i^{\downarrow}$ are the coordinates of spin-up and spin-down.
Most importantly the $\mathbf{K}$ matrix determines the total charge Hall conductance (in units of $e^2/h$) $\sigma_H=\nu=\mathbf{q}^{T}\cdot\mathbf{K}^{-1}\cdot\mathbf{q}$ where $\mathbf{q}$ is the charge vector~\cite{Wen1995}.
Such two-component fractional quantum Hall (FQH) effects emerge as fermionic Halperin (331) states at $\nu=1/2$
in coupled two-layer electron systems~\cite{Yoshioka1988,Yoshioka1989,He1991,He1993}, and further as bosonic Halperin $(221)$ states at $\nu=2/3$ in cold atomic neutral systems~\cite{Grass2012,Furukawa2012,Wu2013,Grass2014}.
Interestingly, the $\mathbf{K}$ matrix and its related Halperin wave function can be generalized to yield a common basis for the description of one-component FQH states~\cite{Regnault2008}, and to construct a large class of symmetry-protected topological phases for multicomponent bosons~\cite{Lu2012,Chen2013}.
In a word, the $\mathbf{K}$ matrix formulism plays an important role
in understanding Abelian multicomponent topological orders.

In contrast, for non-Abelian topological order with non-commutable anyonic braiding statistics, the effective edge theory is known to be more complex than simple free boson theory in Abelian topological orders.
Unfortunately, compared to its success in Abelian topological orders, it is well-known that the non-Abelian topological order cannot be simply characterized by the integer-valued $\mathbf{K}$ matrix as usual.
Despite previous attempts~\cite{Ardonne2000,Lan2019}, there is no well accepted non-Abelian version of the
$\mathbf{K}$ matrix characterization, thus it remains outstanding to precisely describe the internal structure of
non-Abelian topological orders.

Here we explore a way to characterize the internal correlation structure of non-Abelian multicomponent
topological orders based on the Chern-number matrix. So far, it has been successful in characterizing Abelian multicomponent quantum Hall states
based on the Chern number matrix,
 $\mathbf{C}=\frac{1}{m^2-n^2}\begin{pmatrix}
m & -n\\
-n & m\\
\end{pmatrix}$ for $m\neq n$, and $\mathbf{C}$ proved to be the inverse of the $\mathbf{K}=\mathbf{C}^{-1}$ matrix~\cite{Zeng2017,Zeng2018,Zeng2019,Zeng2020,Zeng2022}. Here the diagonal and off-diagonal elements of the Chern-number matrix are related to intracomponent and intercomponent Hall transports respectively.
Nevertheless, it is obscure whether such topological characterization based on the Chern-number matrix can be generalized to the case of non-Abelian multicomponent quantum Hall states.
In view of the physically equivalent relation between the Hall conductance and the Chern number, we aim at the target of constructing the Chern-number matrix for non-Abelian two-component quantum Hall effects.

In this work, we theoretically address the open issues regarding the topological characterization of non-Abelian FQH effects and investigate their Chern-number matrix structures. We will focus on non-Abelian clustered spin-singlet quantum Hall states (dubbed as NASS states)~\cite{Ardonne1999,Reijnders2002,Barkeshli2010}, which are closely theoretical generalizations of the Halperin's wave functions.
In consideration of the NASS wave function,
we conceive a plausible form of the Chern-number matrix to describe the charge responses of the NASS state.
Through state-of-the-art density-matrix renormalization group (DMRG) and exact diagonalization (ED) simulations,
we demonstrate a robust NASS state at filling factor $\nu=4/3$
in topological flat band models with three-body interactions, and then
further validate its quantum Hall responses satisfying the proposed Chern number matrix description.

This paper is organized as follows. In Sec.~\ref{model}, we introduce the SU(2) symmetric Hamiltonian of two-component bosons loaded on two types of topological lattice models, i.e., $\pi$-flux checkerboard and Haldane-honeycomb lattices, and give a description of our numerical methods based on a Chern-number matrix to characterize two-component topological phases. In Sec.~\ref{theory}, we give the theoretical claim for the formalism of the Chern-number matrix of NASS states and connect it to the variational wave function for NASS states. In Sec.~\ref{numeric}, we study the many-body ground states of these two-component bosons in the strong interaction regime, and present numerical results of the Chern-number matrix at the filling factor $\nu=4/3$, based on ED calculation of its Chern number and DMRG calculation of the fractional charge pumping related to Hall conductance. Finally, in Sec.~\ref{summary}, we conclude with a brief discussion of the prospect of generalizing our Chern number matrix to more non-Abelian multicomponent topological phases.

\section{Models and Methods}\label{model}

\begin{figure}[t]
  \centering
  \includegraphics[height=1.6in,width=3.1in]{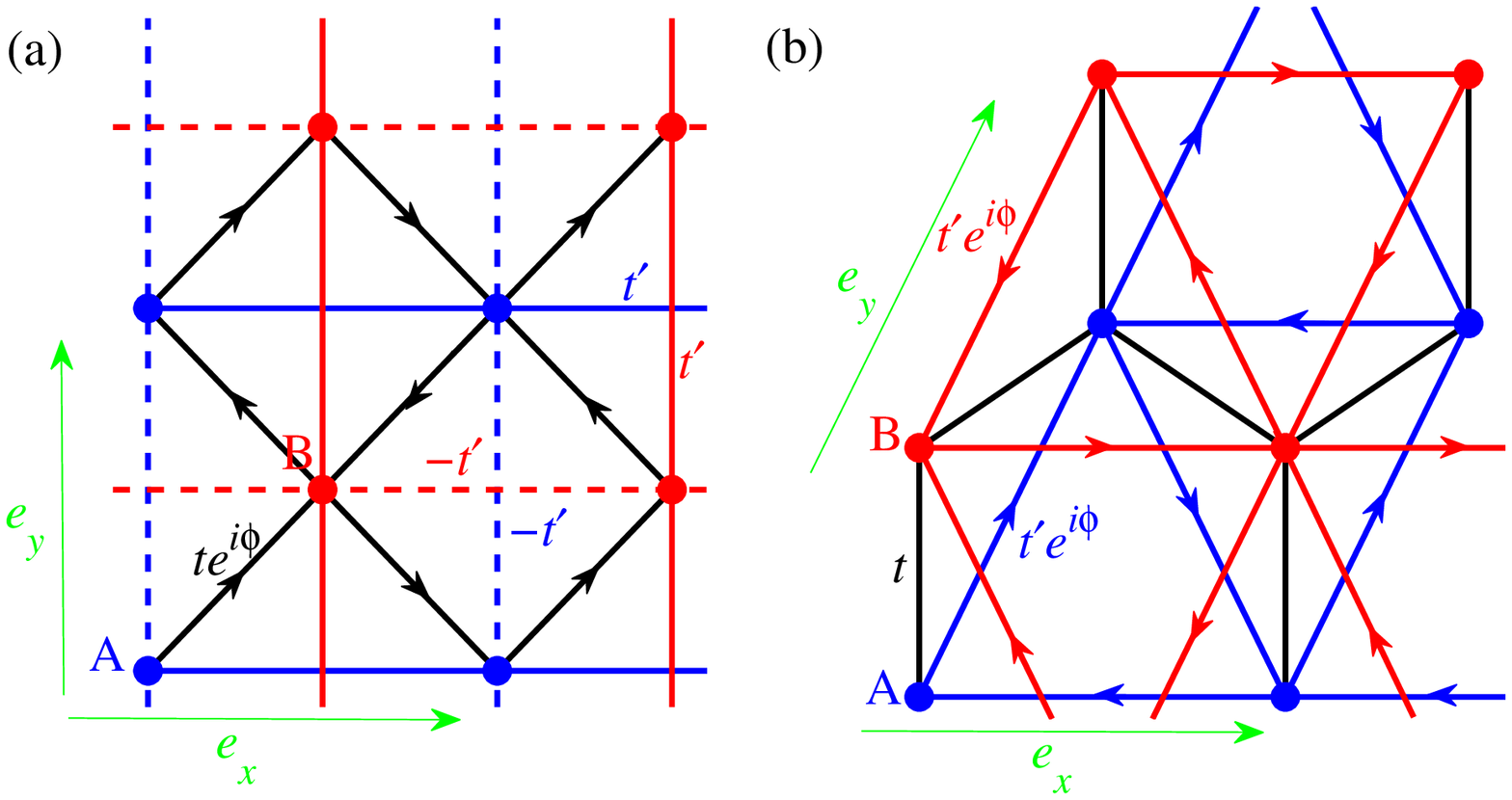}
  \includegraphics[height=1.6in,width=3.1in]{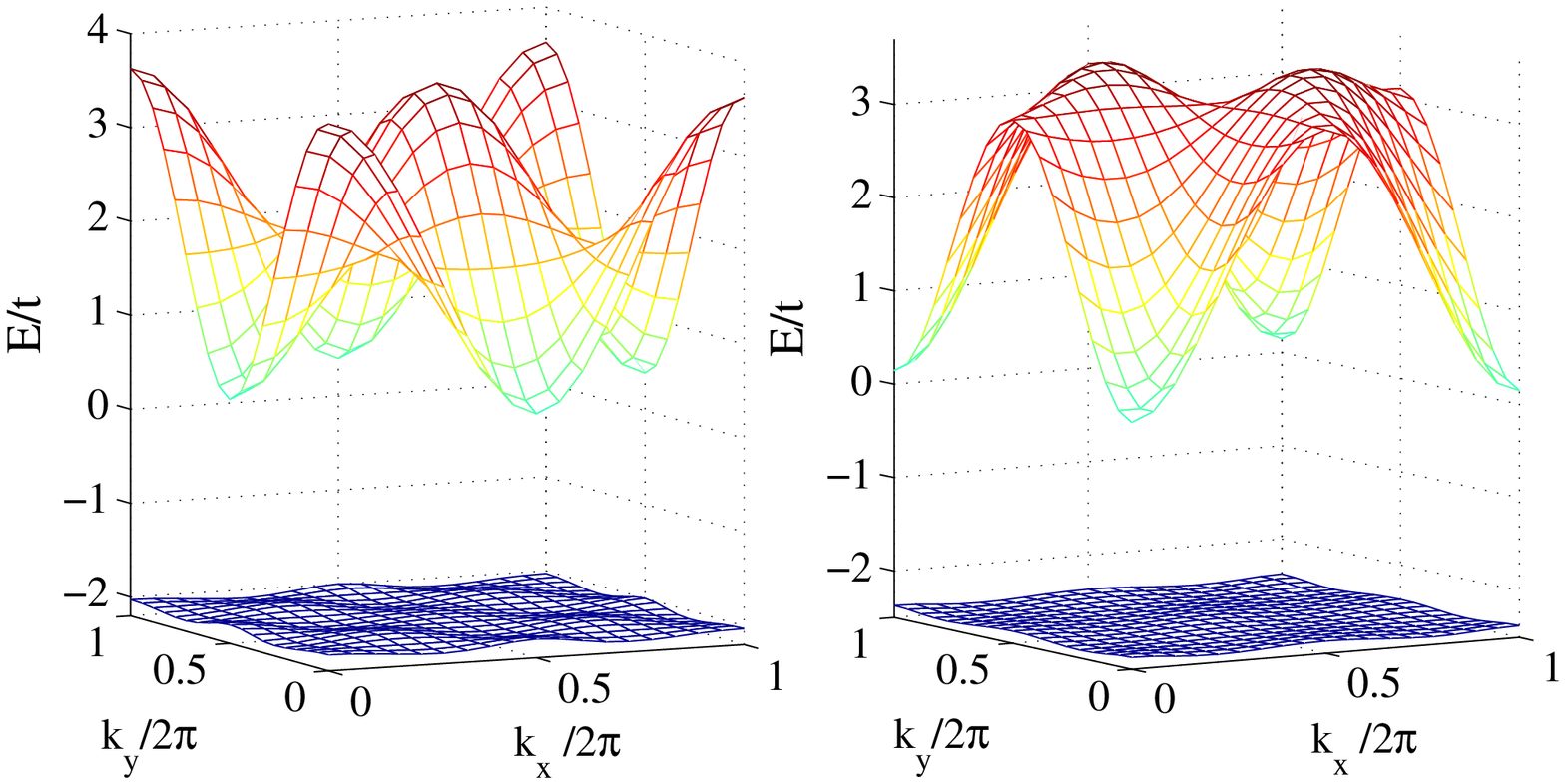}
  \caption{\label{lattice}(Color online) The schematic plot of (a) $\pi$-flux checkerboard lattice model and (b) Haldane-honeycomb lattice model. Two sublattices $A,B$ are labeled by blue (red) filled circles. The arrow link shows the hopping direction carrying a positive sign of chiral flux phase $\phi_{\rr'\rr}=\phi$. For the checkerboard lattice, the next-nearest-neighbor hopping amplitudes are $t_{\rr,\rr'}'=\pm t'$ along the solid (dotted) lines. $e_{x,y}$ indicate the real-space lattice translational vectors. The lower panels show the corresponding band dispersions.}
\end{figure}

Here, we will numerically address the emergence of FQH effect of two-component softcore bosons in topological flat bands through state-of-the-art DMRG and ED simulations, and introduce the Chern-number matrix of two-component systems. We consider the following Hamiltonian for two-component softcore bosons with pseudospin degrees of freedom via $(k+1)$-body interactions at a total filling $\nu=2k/3$ in topological flat bands. The Hamiltonian built on topological $\pi$-flux checkerboard (CB) and Haldane-honeycomb (HC) lattices, is given by
\begin{align}
  H_{CB}=&\!\sum_{\sigma}\!\Big[-t\!\!\sum_{\langle\rr,\rr'\rangle}\!e^{i\phi_{\rr'\rr}}b_{\rr',\sigma}^{\dag}b_{\rr,\sigma}
  -\!\!\!\!\sum_{\langle\langle\rr,\rr'\rangle\rangle}\!\!\!t_{\rr,\rr'}'b_{\rr',\sigma}^{\dag}b_{\rr,\sigma}\nonumber\\
  &-t''\!\!\!\sum_{\langle\langle\langle\rr,\rr'\rangle\rangle\rangle}\!\!\!\! b_{\rr',\sigma}^{\dag}b_{\rr,\sigma}+H.c.\Big]+V_{int},\label{cbl}\\
  H_{HC}=&\!\sum_{\sigma}\!\Big[-t\!\!\sum_{\langle\rr,\rr'\rangle}\!\! b_{\rr',\sigma}^{\dag}b_{\rr,\sigma}-t'\!\!\sum_{\langle\langle\rr,\rr'\rangle\rangle}\!\!e^{i\phi_{\rr'\rr}}b_{\rr',\sigma}^{\dag}b_{\rr,\sigma}\nonumber\\
  &-t''\!\!\sum_{\langle\langle\langle\rr,\rr'\rangle\rangle\rangle}\!\!\!\! b_{\rr',\sigma}^{\dag}b_{\rr,\sigma}+H.c.\Big]+V_{int},\label{hcl}
\end{align}
where $b_{\rr,\sigma}^{\dag}$ is the softcore bosonic creation operator of pseudospin $\sigma=\uparrow,\downarrow$ at site $\rr$, $\langle\ldots\rangle$,$\langle\langle\ldots\rangle\rangle$ and $\langle\langle\langle\ldots\rangle\rangle\rangle$ denote the nearest-neighbor, the next-nearest-neighbor, and the next-next-nearest-neighbor pairs of sites, respectively. We take the $(k+1)$-body Hubbard interaction with pseudospin-SU(2) symmetry,
\begin{align}
  V_{int}=U\sum_{\rr}\prod_{i=0}^{k}(n_{\rr,\uparrow}+n_{\rr,\downarrow}-i)
\end{align}
where $n_{\rr,\sigma}$ is the particle number operator of pseudospin $\sigma$ at site $\rr$, and $U$ is the strength of the onsite interaction. In what follows, we take the tunnel couplings $t'=0.3t,t''=-0.2t,\phi=\pi/4$ for checkerboard lattice, while $t'=0.6t,t''=-0.58t,\phi=2\pi/5$ for honeycomb lattice, such that the lowest Chern band in the whole Brillouin zone is rather flat as indicated in Figs.~\ref{lattice}(a) and~\ref{lattice}(b) for topological lattice geometry, and choose the interaction strength $U=\infty$ (namely no more than $k$ particles are allowed per lattice site).

For finite systems of $N_x\times N_y$ unit cells (enclosing $N_s=2\times N_x\times N_y$ lattice sites), we perform ED calculations on the many-body ground state of the model Hamiltonians Eqs.~\ref{cbl} and~\ref{hcl}. The total filling of the lowest Chern band is $\nu=\nu_{\uparrow}+\nu_{\downarrow}=(N_{\uparrow}+N_{\downarrow})/(N_x\times N_y)$, where $N_{\sigma}$ is the particle number of pseudospin $\sigma$. With the translational symmetry, the energy states are labeled by the total momentum $K=(K_x,K_y)$ in units of $(2\pi/N_x,2\pi/N_y)$ in the Brillouin zone. For larger systems, we exploit the infinite DMRG on the cylindrical geometry, with the maximal bond dimension up to $M=6000$, and the geometry of cylinders is chosen with open boundary condition in the $x$ direction and periodic boundary condition in the $y$ direction.

For the many-body ground state wave function $\psi$ of a given interacting system, the quantized Hall conductance is equivalent to the many-body counterpart of the Chern number~\cite{Niu1985}, which can be calculated using the twisted boundary conditions $\psi(\cdots,\rr_{\sigma}^{i}+N_{\alpha}{\hat e_{\alpha}},\cdots)=\psi(\cdots,\rr_{\sigma}^{i},\cdots)\exp(i\theta_{\sigma}^{\alpha})$
where $\theta_{\sigma}^{\alpha}$ is the twisted angle for particles of pseudospin $\sigma$ in the $\alpha$ direction~\cite{Sheng2003,Sheng2006}. Thus we can build up the Chern-number matrix $\mathbf{C}=\begin{pmatrix}
C_{\uparrow\uparrow} & C_{\uparrow\downarrow} \\
C_{\downarrow\uparrow} & C_{\downarrow\downarrow} \\
\end{pmatrix}$ for a two-component system spanned by pseudospin degree of freedom, with the matrix elements $C_{\sigma\sigma'}=\int d\theta_{\sigma}^{x}d\theta_{\sigma'}^{y}F_{\sigma,\sigma'}(\theta_{\sigma}^{x},\theta_{\sigma'}^{y})/2\pi$ defined in the parameter $(\theta_{\sigma}^{x},\theta_{\sigma'}^{y})$ plane as an integral over the Berry curvature
\begin{align}
    F_{\sigma,\sigma'}^{xy}=\mathbf{Im}\left(\langle{\frac{\partial\psi}{\partial\theta_{\sigma}^x}}|{\frac{\partial\psi}{\partial\theta_{\sigma'}^y}}\rangle
-\langle{\frac{\partial\psi}{\partial\theta_{\sigma'}^y}}|{\frac{\partial\psi}{\partial\theta_{\sigma}^x}}\rangle\right).\nonumber
\end{align}
Within the parameter plane of twisted angles $\theta_{\uparrow}^{x}=\theta_{\downarrow}^{x}=\theta^{x}\subseteq[0,2\pi],\theta_{\uparrow}^{y}=\theta_{\downarrow}^{y}=\theta^{y}\subseteq[0,2\pi]$, we can define the total charge Chern number related to the total charge Hall conductance, as $C_{tot}=\sum_{\sigma,\sigma'}C_{\sigma,\sigma'}=\mathbf{q}\cdot\mathbf{C}\cdot\mathbf{q}^{T}$ where $\mathbf{q}=(1,1)$ is the charge eigenvector of two-component particles. As a concrete example, for $k=1$ in the model Hamiltonians Eqs.~\ref{cbl} and~\ref{hcl}, it is numerically demonstrated that the Chern number matrix $\mathbf{C}_{221}=\frac{1}{3}\begin{pmatrix}
2 & -1\\
-1 & 2\\
\end{pmatrix}$ at $\nu=2/3$ for two-component bosons describes the Halperin (221) states~\cite{Zeng2017}.

\section{Theoretical Scenarios}\label{theory}

In two-component quantum Hall systems, a non-Abelian phase at a given filling $\nu$ could be achieved from two copies of Abelian states at filling $\nu/2$ which are coupled together
by tuning the intercomponent tunneling and/or intercomponent repulsion~\cite{Barkeshli2010a,Barkeshli2010b,Vaezi2014}.
Examples include the non-Abelian Moore-Read $\nu=1$ phase from two bosonic Laughlin $1/2$ FQH states~\cite{Zhu2015},
and non-Abelian phase at $\nu=2/3$ from two fermionic Laughlin $1/3$ FQH states~\cite{Vaezi2014}, which are described by an SU$(N)_1\times$SU$(N)_1\rightarrow$SU$(N)_2$ Chern-Simons-Higgs symmetry-breaking transition mechanism. Across the continuous transition, the final Chern number of the emerging non-Abelian state is given by $\sigma_H=\nu=\nu/2+\nu/2$, which is the sum of the Chern numbers of two identical Abelian states.

Alternatively, Ref.~\cite{Ardonne1999} proposed a series of non-Abelian spin-singlet quantum Hall phases at filling factors $\nu=2k/(2kM+3)$ with spinful SU(2) symmetry (labeled by a pseudospin quantum number).
Here, we claim that the NASS quantum Hall states can be identified by a class of Chern number matrices
\begin{align}
  \mathbf{C}\!&=\!\begin{pmatrix}
C_{\uparrow\uparrow} & C_{\uparrow\downarrow}\\
C_{\downarrow\uparrow} & C_{\downarrow\downarrow}\\
\end{pmatrix}\!\nonumber\\
&=\!\frac{k}{2kM+3}\!\begin{pmatrix}
kM+2 & -(kM+1)\\
-(kM+1) & kM+2\\
\end{pmatrix}\label{chernmatrix}
\end{align}
where even $M$ for bosons and odd $M$ for fermions. This form of the Chern number matrix satisfies the following two sufficient conditions.
First, we recover the total charge Chern number as $C_{tot}=\sum_{\sigma,\sigma'}C_{\sigma\sigma'}=\nu$ which is exactly conserved to the filling factor, thus topological invariant $C_{tot}$ is connected to the total Hall conductance, as it should be.
Second, akin to the Halperin state, the Chern number matrix elements determine the strength of the intracomponent and intercomponent
correlations respectively through the variational many-body wave function (apart from a constant conformal product factor)~\cite{Ardonne1999}:
\begin{align}
 \Psi_{\nu} \propto \prod_{i<j}(z_i^{\uparrow}-z_j^{\uparrow})^{l_{\uparrow\uparrow}} (w_i^{\downarrow}-w_j^{\downarrow})^{l_{\downarrow\downarrow}} \prod_{i,j}(z_i^{\uparrow}-w_j^{\downarrow})^{l_{\uparrow\downarrow}},\nonumber
\end{align}
where $z_i^{\uparrow},w_i^{\downarrow}$ are the coordinates of spin-up and spin-down and the rational exponents $l_{\uparrow\uparrow}=l_{\downarrow\downarrow}$ and $l_{\uparrow\downarrow}=l_{\downarrow\uparrow}$ are recognized to meet the relation
\begin{align}
  \begin{pmatrix}
l_{\uparrow\uparrow} & l_{\uparrow\downarrow}\\
l_{\downarrow\uparrow} & l_{\downarrow\downarrow}\\
\end{pmatrix}=\mathbf{C}^{-1}=\frac{1}{k}\begin{pmatrix}
kM+2 & kM+1\\
kM+1 & kM+2\\
\end{pmatrix}.
\end{align}

In effective conformal field theory for NASS, in the simplest case $M=0$ where the SU$(3)_k$ symmetry is preserved [for $M\neq0$ the SU$(3)_k$ symmetry is broken], the SU$(3)_k$ field theory can be constructed by pairing $k$ copies of the SU$(3)_1$ field theory of Abelian Halperin (221) state together.
For instance, let us consider the case of SU$(3)_2$ NASS state at filling $\nu=4/3$ by coupling two SU$(3)_1$ Halperin $(221)$ states at filling $\nu=2/3$ with a symmetry breaking SU$(3)_1\times$SU$(3)_1\rightarrow$SU$(3)_2$.
In consideration of the Chern-number matrix $\mathbf{C}_{221}=\frac{1}{3}\begin{pmatrix}
2 & -1\\
-1 & 2\\
\end{pmatrix}$ of Halperin (221) states~\cite{Zeng2017}, the associated matrix from Eq.~\ref{chernmatrix} shows the coincidence $\mathbf{C}_{\nu=4/3}=\mathbf{C}_{221}+\mathbf{C}_{221} = 2\times\mathbf{C}_{221}$, similar to the above case of the Moore-Read state obtained from two coupled Abelian states. More generally, we can construct the Chern-number matrix $\mathbf{C}_{\nu=2k/3}=k\times\mathbf{C}_{221}$ of SU$(3)_k$ NASS quantum Hall states at $\nu=2k/3\quad(k>2)$ as the integer $k$ multiples of $\mathbf{C}_{221}$, with the symmetry breaking SU$(3)_1\times$SU$(3)_1\times\cdots$SU$(3)_1\rightarrow$SU$(3)_k$. In this context, we believe the Chern-number matrix provides a faithful description of charged quantum Hall responses of NASS state. For $k>1$, however, to our best knowledge the study of their Chern number matrices in microscopic lattice systems is still lacking, which is the focus of the present work. We shall elucidate the characteristic topological degeneracy, topologically invariant Chern number, and fractional charge pumping, to dissect the topological information of two-component systems.

\begin{figure}[t]
  \includegraphics[height=1.9in,width=3.4in]{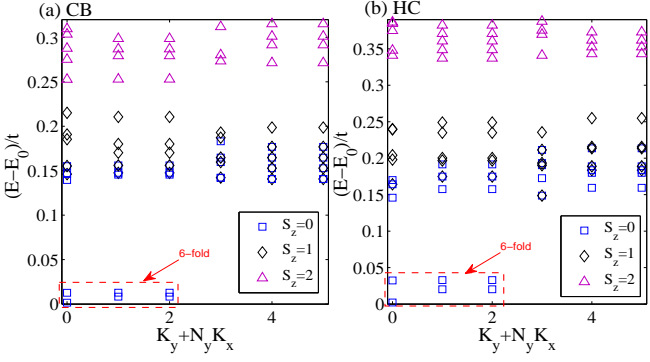}
  \caption{\label{energy} (Color online) Low energy spectrum of two-component bosonic systems $N_{\uparrow}+N_{\downarrow}=8,N_s=2\times2\times3$ with different spin sectors $S_z$ for infinite three-body interaction in different topological lattices: (a) $\pi$-flux checkerboard lattice and (b) Haldane-honeycomb lattice.}
\end{figure}

\section{Numerical Results for Chern-number matrix}\label{numeric}

Following the last section, we begin to systematically present numerical results for the topological properties of many body ground states at $\nu=4/3$ with infinite three-body interaction $(k=2)$ in the model Hamiltonians Eqs.~\ref{cbl} and~\ref{hcl}. For $(k+1)$-body interaction, the ground state degeneracy is $(k+1)(k+2)/2$ for non-Abelian spin-singlet FQH states at $\nu=2k/3$. In the ED study of finite system sizes, we plot the low-energy spectrum for different topological lattice models as shown in Figs.~\ref{energy}(a) and~\ref{energy}(b). We find that the ground states host a six-fold degeneracy with a robust protecting gap separated from higher energy levels. These six-fold degenerate ground energies $E_0(S_z)$ always fall into the total spin sector $S_z=(N_{\uparrow}-N_{\downarrow})/2=0$, and they satisfy with $E_0(S_z)<E_0(S_z+1),S_z\geq0$, in consistency with spin-singlet nature.

\begin{figure}[t]
  \includegraphics[height=2.5in,width=3.4in]{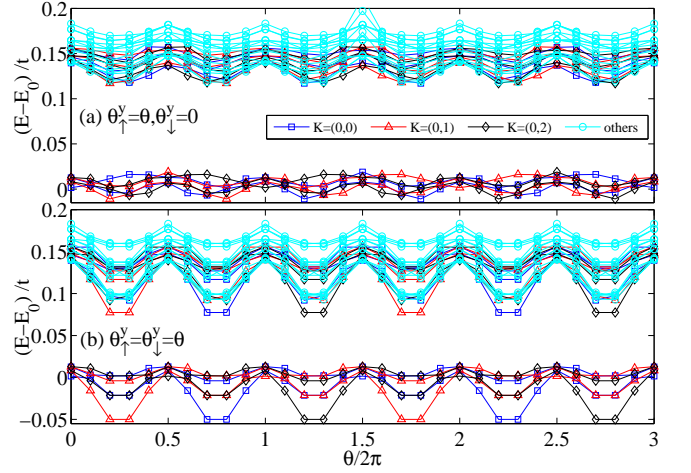}
  \caption{\label{flux} (Color online) Low energy spectral flow of two-component bosonic systems $N_{\uparrow}=N_{\downarrow}=4,N_s=2\times2\times3$ for infinite three-body interaction in topological checkerboard lattice under the insertion of two different types of flux quanta (a) $\theta_{\uparrow}^{y}=\theta,\theta_{\downarrow}^{y}=0$ and (b) $\theta_{\uparrow}^{y}=\theta_{\downarrow}^{y}=\theta$.}
\end{figure}

\begin{figure}[b]
  \includegraphics[height=1.58in,width=3.4in]{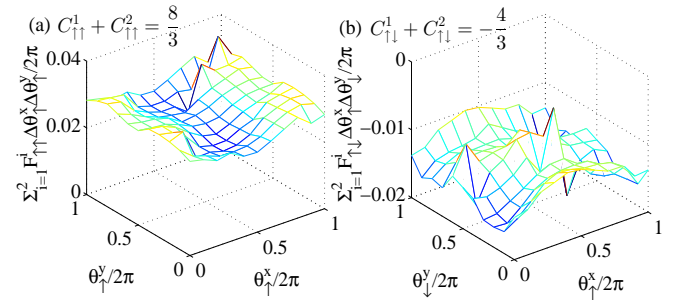}
  \caption{\label{berry} (Color online) Distribution of Berry curvatures $F_{\sigma,\sigma'}^{xy}\Delta\theta_{\sigma}^{x}\Delta\theta_{\sigma'}^{y}/2\pi$ of the two ground states at momentum $K=(0,0)$ of two-component bosonic systems $N_{\uparrow}=N_{\downarrow}=4,N_s=2\times2\times3$ for infinite three-body interaction in topological checkerboard lattice under different twisted angles: (a) $(\theta_{\uparrow}^{x},\theta_{\uparrow}^{y})$ with $\theta_{\downarrow}^{x}=0,\theta_{\downarrow}^{y}=0$ and (b) $(\theta_{\uparrow}^{x},\theta_{\downarrow}^{y})$ with $\theta_{\downarrow}^{x}=0,\theta_{\uparrow}^{y}=0$.}
\end{figure}

Further, in order to demonstrate the robustness of these topologically degenerate ground manifolds,
we calculate the low energy spectra flux under the insertion of flux quanta $\theta_{\sigma}^{\alpha}$ ($\alpha=x,y$). As indicated in Figs.~\ref{flux}(a) and~\ref{flux}(b), these six-fold ground states at different momentum sectors $K=(0,i) (i=0,1,2)$ are shifted into each other without mixing with the higher energy levels,
and the system goes back to itself upon the insertion of three flux quanta for both $\theta_{\uparrow}^{\alpha}=\theta_{\downarrow}^{\alpha}=\theta$
and $\theta_{\uparrow}^{\alpha}=\theta,\theta_{\downarrow}^{\alpha}=0$, implying the fractional quantization of topological invariant.

Next we extract the Chern-number matrix $\mathbf{C}$ by numerically calculating the Berry curvatures using $m\times m$ mesh squares in the boundary phase space with $m\geq10$. For the two ground states at momentum $K=(0,0)$ of two-component bosonic system $N_{\uparrow}=N_{\downarrow}=4,N_s=2\times2\times3$, we find that intracomponent Berry curvatures $F_{\uparrow\uparrow}^{xy}$ have an opposite sign to that of intercomponent Berry curvatures $F_{\uparrow\downarrow}^{xy}$, and obtain the quantized Chern numbers $\sum_{i=1}^2C^i_{\uparrow\uparrow}=8/3$ and $\sum_{i=1}^2C^i_{\uparrow\downarrow}=-4/3$ with the symmetric relation $C_{\uparrow\uparrow}=C_{\downarrow\downarrow},C_{\uparrow\downarrow}=C_{\downarrow\uparrow}$, as indicated in Figs.~\ref{berry}(a) and~\ref{berry}(b) respectively. Similarly, for six quasidegenerate ground states at momentum sectors $K=(0,i) (i=0,1,2)$, we obtain the relationship $\sum_{i=1}^6C^i_{\uparrow\uparrow}=8$ and $\sum_{i=1}^6C^i_{\uparrow\downarrow}=-4$. The above results imply a symmetric Chern-number matrix
\begin{align}
  \mathbf{C}=\begin{pmatrix}
C_{\uparrow\uparrow} & C_{\uparrow\downarrow}\\
C_{\downarrow\uparrow} & C_{\downarrow\downarrow}\\
\end{pmatrix}=\frac{1}{3}\begin{pmatrix}
4 & -2\\
-2 & 4\\
\end{pmatrix}\label{chern}
\end{align}
It is exactly the same with the case of $M=0,k=2$ in Eq.~\ref{chernmatrix}.

\begin{figure}[t]
  \includegraphics[height=1.8in,width=3.4in]{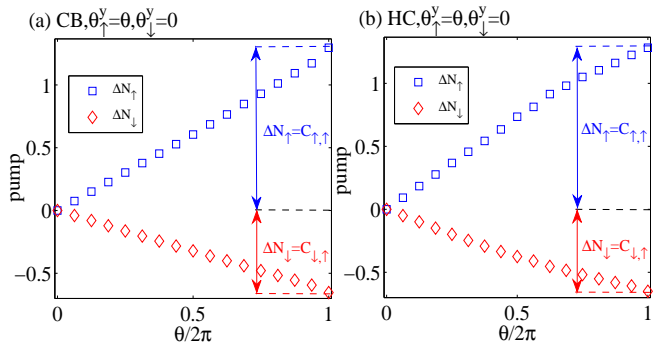}
  \caption{\label{pump} (Color online) Quantized charge transfers of two-component bosonic systems $\nu_{\uparrow}=\nu_{\downarrow}=2/3$ for infinite three-body interaction under the insertion of flux quantum $\theta_{\uparrow}^{y}=\theta,\theta_{\downarrow}^{y}=0$ on an infinite $N_y=3$ cylinder of different topological lattices: (a) $\pi$-flux checkerboard lattice and (b) Haldane-honeycomb lattice. }
\end{figure}

For larger system sizes, we further carry out the charge pumping under the insertion of a flux quantum $\theta_{\sigma}^{y}$ on infinite cylinder systems using the DMRG~\cite{Gong2014}. In the parameter period $\theta_{\sigma}^{y}\subseteq[0,2\pi]$, due to the Berry curvature the particles of pseudospin $\sigma'$ is pumped along the $x$ direction determined by the topological invariant $C_{\sigma',\sigma}$ in connection with the quantized Hall conductance. Numerically we cut the cylinder along the $x$ direction into two halves. The charge increment of the particles of pseudospin $\sigma'$ on the left part of infinite cylinder systems is encoded by the particle number of pseudospin $\sigma$ on the left cylinder part $N_{\sigma'}(\theta_{\sigma}^{y})=tr[\widehat{\rho}_L(\theta_{\sigma}^{y})\widehat{N}_{\sigma'}]$ (here $\widehat{\rho}_L$ is the reduced density matrix of the left part). At $\nu_{\uparrow}=\nu_{\downarrow}=2/3$, as shown in Figs.~\ref{pump}(a) and~\ref{pump}(b) for different topological lattice models, we obtain the universal charge transfers of the particles of pseudospin $\sigma=\uparrow,\downarrow$ with fractionally quantized pumping values
\begin{align}
  &\Delta N_{\uparrow}=N_{\uparrow}(\theta_{\uparrow}^y=2\pi)-N_{\uparrow}(\theta_{\uparrow}^y=0)\simeq C_{\uparrow\uparrow}=\frac{4}{3},\nonumber\\
  &\Delta N_{\downarrow}=N_{\downarrow}(\theta_{\uparrow}^y=2\pi)-N_{\downarrow}(\theta_{\uparrow}^y=0)\simeq C_{\downarrow\uparrow}=-\frac{2}{3} \nonumber
\end{align}
within one flux quantum period from $\theta_{\uparrow}^y=0$ to $\theta_{\uparrow}^y=2\pi$ with $\theta_{\downarrow}^y=0$ for two-component bosons. Thus our DMRG study, in line with the ED study, confirms the formulation of the Chern-number matrix for the $\nu=4/3$ NASS state.

\section{Conclusion and Outlook}\label{summary}

To summarize, we have opened up a particular sketch of the Chern-number matrix given by Eq.~\ref{chernmatrix} to
describe the multicomponent non-Abelian spin-singlet quantum Hall states, and numerically verified it using two-component interacting bosons on topological lattice models.
We study a microscopic model of two-component interacting bosons in two typical topological lattices
which can realize non-Abelian spin-singlet FQH states at a commensurate partial filling $\nu=4/3$ in the lowest Chern band with tailored three-body interactions.
We have proved numerically that the many-body ground states are topologically characterized by the six-fold degenerate manifold. Crucially, the fractional quantized charge pumping is faithfully consistent with the elements in Chern number matrix.
We anticipate that the universal formulation of the Chern-number matrix holds for sequential fillings $\nu=2k/3$ with $k>2$.

\begin{figure}[t]
  \includegraphics[height=1.72in,width=3.4in]{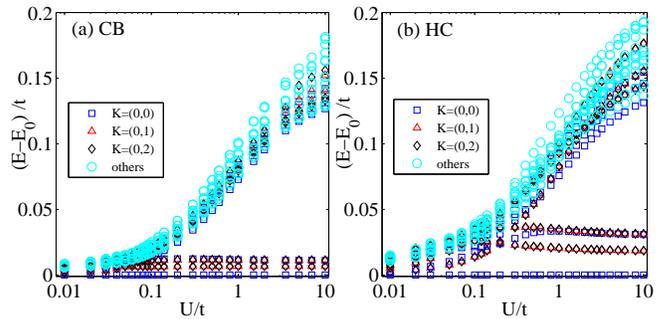}
  \caption{\label{intU} (Color online) Numerical ED results for the low energy spectrum of two-component bosonic systems $N_{\uparrow}=N_{\downarrow}=4,N_s=2\times2\times3$ in different topological lattices: (a) $\pi$-flux checkerboard lattice and (b) Haldane-honeycomb lattice, as finite onsite three-body interaction $U$ is tuned from weak repulsion $U/t=0.01$ to strong repulsion $U/t=10$. Only the lowest five energy states in each momentum sector are shown.}
\end{figure}

Finally, several remarks are given in order. Physically, the Chern number matrix Eq.~\ref{chernmatrix}
reveals the charge response of the NASS state, which is experimentally accessible in transport measurements directly using electric Hall conductance, while the exotic pairing neutral modes are beyond the scope of the above description. A more general paradigm capturing all non-Abelian features deserves exploration.
Each element $C_{\sigma,\sigma'}$ in the Chern number matrix determines either the
intracomponent Hall transport (when $\sigma=\sigma'$) or the intercomponent drag Hall transport (when $\sigma\neq\sigma'$), as demonstrated for interlayer quantum Hall effect in a coupled graphene double layer~\cite{Li2019,Liu2019}.
Nevertheless the present work based on the Chern number matrix $\mathbf{C}$ would provide us a possible route for the identification of other nontrivial non-Abelian two-component FQH states through the relation $\begin{pmatrix}
l_{\uparrow\uparrow} & l_{\uparrow\downarrow}\\
l_{\downarrow\uparrow} & l_{\downarrow\downarrow}\\
\end{pmatrix}=\mathbf{C}^{-1}$, such as possible fermionic non-Abelian $\nu=4/5,4/7,4/9$ states in bilayer electronic systems~\cite{Barkeshli2010b}, with the similar wave function structure
\begin{align}
 \Psi_{\nu} \propto \prod_{i<j}(z_i^{\uparrow}-z_j^{\uparrow})^{l_{\uparrow\uparrow}} (w_i^{\downarrow}-w_j^{\downarrow})^{l_{\downarrow\downarrow}} \prod_{i,j}(z_i^{\uparrow}-w_j^{\downarrow})^{l_{\uparrow\downarrow}}.\nonumber
\end{align}
In conclusion,
this current work also shows that our two-component flat band model is a fascinating playground to realize bosonic non-Abelian spin-singlet phases which may be testable for cold atom experiments~\cite{Cooper2019}.

\begin{acknowledgements}
T.S.Z. particularly thanks D. N. Sheng for inspiring discussions on multicomponent quantum Hall effects and careful readings of our manuscript. This work is supported by the National Natural Science Foundation of China (NSFC) under Grant No. 12074320 (T.S.Z.), and Grant Nos. 11974288, 92165102 (W.Z.). This work was also supported by the ``Pioneer'' and ``Leading Goose'' R\&D Program of Zhejiang Province (2022SDXHDX0005) and the foundation from Westlake University.
\end{acknowledgements}

\appendix

\section{Finite interaction effects}

Contrast to infinite three-body repulsion $U\rightarrow\infty$, here we consider the effect of finite softening repulsion $U$ on the low energy spectrum of the model Hamiltonians Eq.~\ref{cbl} and Eq.~\ref{hcl}. In Figure~\ref{intU}, we plot the low energy spectrum as a function of $U$ for different topological lattices. For very weak repulsions $U/t\ll1$, bosons in each layer are nearly decoupled to each other, and the low energy spectrum hosts a large dense energy density, implying a gapless compressible liquid. However for very strong repulsions $U/t\gg1$, a robust ground degenerate manifold is preserved. Our calculation of different model systems of a finite size gives a small critical repulsion $U_c\simeq0.1t$, in order to maintain six-fold topological degeneracy. Actually this critical value $U_c$ is determined by looking into the point where the six-fold degeneracy disappears when these ground states begin to mix with excited energy levels.

\end{document}